%Paper: hep-ph/9207210
%From: MILLER@alpher.npl.washington.edu
%Date: Fri, 3 Jul 1992 8:15:19 PDT

% This is a TEX file
%10 ps files are available directly from the author

\normalbaselineskip=20pt
\baselineskip=20pt
\magnification=1200
\overfullrule=0pt
\hsize 16.0true cm
\vsize 22.0true cm

\def\lam{{ \Lambda^2 }}

\def\phat{{\widehat p}}
\def\Mhat{{\widehat M}}

\def\U{{\widehat U}}
\def\V{{\widehat V}}
\def\lambar{{\overline{\lambda}}}
\def\bop2{{ {{ {\widehat b}^2}\over {b_H^2}} }}
\def\B {{ \vec B }}

\def\so2{{ {{\sigma}\over 2} }}
\def\iout#1#2{{ \int_{-\infty}^{#1} d#2 \rho(\B,#2) }}
\hfill 40427-23-N92, hep-ph/9207210
\centerline{{\bf  Multiple-Scattering Series For Color Transparency}}
\vskip 18pt
\centerline {W. R. Greenberg and G. A. Miller}
\vskip 18pt
\centerline {\it Department of Physics, FM-15}
\centerline {\it University of Washington, Seattle WA. 98195}
\vskip 24pt
\centerline{\bf Abstract}
Color transparency CT depends on the formation of a wavepacket of small
spatial extent. It is useful to interpret
 experimental searches for CT  with
a multiple scattering scattering series based on wavepacket-nucleon
scattering instead of the standard one using nucleon-nucleon scattering.
We develop several new techniques which are valid for differing ranges of
energy. These techniques are applied to verify some early approximations;
study new forms of the wave-packet-nucleon interaction;
examine effects of treating wave packets of non-zero size; and predict the
production of $N^*$'s in electron scattering experiments.
\vskip 48pt
\centerline{Submitted to: {\it Physical Review D}}
\vfill \eject

\centerline{I.  INTRODUCTION}
\nobreak
Color Transparency (CT) is the postulated [1,2]
absence of final (or initial) state interactions
caused by the cancellation of color fields of a system of quarks and gluons
with small spatial separation.  For example, suppose an electron impinges on a
nucleus knocking out a proton at high momentum transfer.  The consequence of
color transparency is that there is no exponential loss of flux as the ejected
particle propagates through the nucleus.  Thus, the usually ``black" nucleus
becomes transparent. We restrict our attention to
processes for which the fundamental reaction is elastic, or at least a
two-body reaction.  This requires that the  nuclear excitation energy
be known well enough
to ensure that no extra pions are created.
This subject is under active experimental investigation [3-5].

  The existence of color transparency depends on: (1) the formation of a
small-sized wave-packet in a high momentum transfer reaction. (2) the
interaction between such a small object and nucleons being suppressed (color
neutrality or screening) and (3) the wave-packet escaping the nucleus while
still small [1,2].
That color neutrality (screening)
causes the cross section of small-sized color singlet
configurations with hadrons to be small was found in Ref.~[6], and is
well-reviewed in
Refs. [7-9].  So we take item (2) as given. The truth of item (1),
for experimentally available energies,
is an interesting issue. It is discussed in Refs. [9-11], and is not probed
in depth here.

It is also true that  at
experimentally available energies, the small object does expand as it moves
through the nucleus. Thus the final state interactions are suppressed but not
zero. The importance of this expansion was found by Farrar et al.~[12], and by
Jennings and Miller~[13]. See also Ref.~[14].
The purpose of this paper is to determine improved methods to calculate
the evolution of a wave packet as it moves through a nucleus.
In Ref.~[12] the uncertainty principle is  used
to argue that the wave-packet-nucleon interaction can be treated by using an
effective cross section, $\sigma_{eff}(Z)\propto Z$ where $Z$ is the
propagation length.  In Ref.~[13], the wave-packet-nucleon interactions are
computed using a hadronic basis.
An approximation to the full multiple scattering series is made in which
the first-order term is exponentiated. The work of Ref.~[14] uses an exact
Green's function for the evolution of the wave-packet in a
model in which the wave-packet-nucleon interaction is
proportional to the sum of the squares of the transverse separations
between the quarks in the ejected wave-packet, $b^2$.  Moreover, the
baryon states are treated as quarks bound in a two-dimensional harmonic
oscillator. The exact solution depends on using this particular
interaction and model space.

The aim of the present work is to develop techniques and approximations
that allow the use of more general
models of the wavepacket-nucleon interaction
and baryon space. A general formal expression for the necessary
multiple scattering series is derived in Section II. Different approximation
schemes
are also defined in that section. Numerical studies of the different
approximations are made, using the $(e,e'p)$ reaction
as an example, in Sect III. (The initial wave packet is assumed to be of
zero size.) Approximations that work best in different
regions of energy are identified.
The applications in Sect. III result from using a wave-packet-nucleon
interaction proportional to $b^2$. Sect. IV shows how different interactions,
e.g. that of Ref.~[15], can be used. This
demonstrates the general nature of our methods.
Sect. V displays how our techniques are used
in situations in which the wavepacket is allowed to
have a non-zero size. Another application is to $N^*$ production, Sect. VI.
The final section is reserved  for a few brief summary remarks.
\bigskip
\centerline{II.  FORMALISM AND APPROXIMATIONS}
\nobreak
Our technique is to treat the wavepacket, formed
in high momentum transfer processes, as a coherent sum of baryon states.
The means that the relevant wave equation should be a matrix equation.
Hence
the starting point for our analysis is the relativistic wave
equation
$$\left(\nabla^2 + \phat^2 \right) | \Psi_{N,p} \rangle = U | \Psi_{N,p}
\rangle.
\eqno(2.1)$$
The quantity $|\Psi_{N,p}\rangle$ represents a vector in internal quark and
nuclear
(center of mass position) spaces.  The subscripts refer to the asymptotic
boundary condition that a free nucleon (N) has a momentum $\vec p$.
  The operator
$\phat$ is
$$\phat = \sqrt{p^2 + M^2 - \Mhat^2},\eqno(2.2)$$
$$\approx p + {M^2 - \Mhat^2 \over 2p}\,, \eqno(2.3)$$
where the energy eigenstates of the quark ``Hamiltonian", $\Mhat^2$, are
such that
$$\Mhat^2 | n \rangle = | n \rangle M^2_n. \eqno(2.4)$$
In this notation, the nucleon is the ground state, with n = N with eigenvalue
$M^2$. The matrix operator $U$ acts in both quark and nuclear space. It has
matrix
elements between the nucleon and its excited states and also depends on the
nuclear density $\rho(\vec R)$.

We wish to simplify Eq.~(2.1). To this end, let $\vec R$ be the operator
denoting
the position of the center of mass of the struck nucleon.
Also, we take the final nucleon momentum, $\vec p$, to be large and in the
$\widehat Z$ direction.  Define another
column vector ${\bf \Psi}_p (\vec R)$ as
$${\bf \Psi}_p (\vec R) \equiv \langle \vec R | \Psi_{N,p} \rangle\,.\eqno(2.5)
$$
It is convenient to factor out a plane wave factor and write,
$${\bf \Psi}_p (\vec R) = e^{i \phat Z} \, \widehat  \varphi (\vec R) | N
\rangle. \eqno(2.6)$$
This is the defining equation for $\widehat \varphi (\vec R)$, a matrix in
baryon
space.  Then, as an operator equation, assuming the momentum is large compared
with any gradient in the problem,
$$2i \phat \, {\partial \widehat \varphi (\vec R) \over \partial Z}  = \U \,
(\vec R) \, \widehat \varphi \, (\vec R), \eqno(2.7)$$
where
$$\U \, (\vec R) \equiv \, e^{-i \phat Z} \, U (\vec R) \, e^{i \phat
 Z} \,. \eqno(2.8)$$
The solution to Eq.~(2.7) is a path ordered exponential,
$$\widehat \varphi (\vec R) \, = \, {\cal P} \, \exp \left(\int^Z_{- \infty}
\, {d Z_1
\over 2i\phat} \U (\vec B, Z_1) \right) \,. \eqno(2.9)$$
Here, $\cal P$ is the path ordering symbol.  Notice that we have chosen to use
outgoing boundary conditions and that $\vec R = \vec B + Z \widehat Z$.  We
often suppress the index $\vec B$ in evaluating the path integral.

Before proceeding it is worthwhile to discuss the nuclear interaction operator
$U (\vec R)$.  This is the product of the wave-packet-nucleon interaction $4
\pi \widehat f$ by the target nuclear density $\rho (\vec R)$
$$U \, (\vec R) = 4 \pi \widehat f \rho (\vec R). \eqno(2.10)$$
In the forward scattering approximation used here, the operator $\widehat f$
acts on the internal quark space only.  We write $\widehat f = f (b^2)
$ where $b_{ij}$ is the transverse (perpendicular to $\vec p$)
separation operator between quarks i and j in the wave
packet and $b^2
\equiv \sum \limits_{ij} b^2_{ij}$.
That the interaction does not depend on the longitudinal coordinates is
suggested by the work of Ref.~[6]; see also Ref.~[8]. Although
there is no detailed theory for $\widehat f$,
some general things are known.  For small wave packets with $b^2 \ll
b^2_H$ where $b^2_H = \langle N | b^2 | N \rangle$, $\widehat f$ should vanish.
Interactions do occur for larger wavepackets. The expectation value of
$\widehat f$
in a normal baryon should be the experimentally measured forward scattering
amplitude.  Such are dominated by their imaginary part so we keep only $Im
\widehat f$.  Thus we write
$$4 \pi \widehat f \, = \, -i \, 2 \phat {\sigma \over 2}\,f (b^2),
\eqno(2.11a)$$
where $\sigma$ is the proton-nucleon total cross section with
$$f(b^2 = 0) = 0, \eqno(2.11b)$$
$$\langle N | f (b^2) | N \rangle = 1. \eqno(2.11c)$$
In the two gluon exchange approximation,
$$\lim \limits_{b^2 \rightarrow 0} \, f(b^2) \, \approx {b^2 \over b^2_H}.
\eqno(2.11d)$$
We shall use a general function $f (b^2)$, subject only to the constraints of
Eqs.~(2.11).  The presence of the factor $\phat$ allows the high
energy baryon-nuclear cross section to be independent of the beam energy.
Replacing $\phat$ by $p$ in Eq.~(2.8) is of the same order of approximation
as those already made and produces no essential changes in our results.

To be specific, we examine the $(e, e' p)$
reaction~[4] and compute amplitudes for processes in which a proton is
knocked out of some shell model bound state, $\alpha$.  With
semi-exclusive kinematics, as in the experiment done at SLAC, we look for an
exiting proton which has momentum, $\vec p$, very close to that of the virtual
photon, whose momentum we label by $\vec q$.  By close we mean that no extra
pions are produced in the collision.  Then, we define the ``knockout"
amplitude and cross section by the overlap
$${\cal M}_\alpha = \langle N , \alpha | T_H (Q^2) | \Psi _{N,p} \rangle ,
\eqno(2.12a)$$
$$\sigma \propto \sum \limits_\alpha |M_\alpha|^2,\eqno(2.12b)$$
where the sum is over all the occupied shells $\alpha$.
The cross section of Eq.~(2.12) is obtained by integrating over the angle of
the outgoing proton, corresponding to the SLAC experiment.  We shall be
concerned with ratios $\sigma/\sigma^B$ in which $\sigma^B$ is obtained by
replacing $\langle \vec R |\Psi_{N,p} \rangle$ of Eq.~(2.12), as given by
Eqs.~(2.5), (2.6), and (2.9), by a proton plane
wave function.

The operator $T_H (Q^2)$ accounts for the absorption of a high momentum photon
by a nucleon.  Ignoring the effects of spin, the matrix elements of the hard
scattering operator are
$$\langle \vec R, N | \, T_H (Q^2) | \vec R, m \rangle = e^{-i q Z} \,
\delta^{(3)} \, (\vec R - \vec R') F_{N,m} (Q^2), \eqno(2.13)$$
where $F_{N,m} (Q^2)$ is the inelastic transition form factor.  The term
$e^{-i q Z} \sum \limits_m \, F_{N,m} (Q^2) \langle m |$ represents a high
momentum wave packet propagating through the nucleus.

After inserting a complete set of nuclear position states, the scattering
amplitude ${\cal M}_\alpha$ has the expression:
$${\cal M}_\alpha = \int d^3 R \,\Phi_\alpha (\vec R)\, \langle N | T_H (Q^2)
\, e^{i \phat
Z} {\cal P} \, \exp \left(\int^Z_{- \infty} \, d Z_1 \widehat V (\vec B, Z_1)
\right) | N \rangle, \eqno(2.14)$$
where $\Phi _\alpha (\vec R)$ is the wavefunction of a nucleon bound to
the nucleus in shell $\alpha$, and $\vec q$ is the three momentum of the
incoming photon.  We take the shell model parameter $\hbar \omega = 41 \,
MeV $.  The operator $\widehat V$ is given by
$$\widehat V (\vec R)\equiv {1\over 2i \phat }\U (\vec R),\eqno(2.15a)$$
$$\widehat V (\vec R)= - e^{-i \phat Z} \, {\sigma \over 2} \, \rho (\vec R) \,
f (b^2) \, e ^{i \phat Z}. \eqno(2.15b)$$
The ``hat'' over $V$ is meant to symbolize that $\widehat V (\vec R)$ also
depends on $b^2$.
We note that in general, $\left[ \V (\vec R), \, \V (\vec R')\right] \neq 0$
since
the operators $\phat$ and $b^2$ do not commute.
All of the nuclear
interactions are contained in the scattering wave function $| \Psi_{N,p}
\rangle.$  At initial stages the wave packet is small and so are interactions.
If the wave packet expands, the influence of interactions is more important.
Thus, an accurate approximation must apply for both small and large sized wave
packets.  The aim of this work is to develop approximation techniques for
Eq.~(2.9) or Eq.~(2.14).
This involves handling the evolution of an initially small wave packet as
it moves through the nucleus.

At this stage there are several things that one can do to obtain $\widehat
\varphi$ and ${\bf \Psi}_p$.  We will consider
what we call the ``Order By Order'' calculation, the ``Low Energy
Expansion'', then the ``High Energy Expansion''
and the ``Equal Spaced Momentum''
replacement, and finally
the ``Exponential Approximation''.
\medskip
\centerline{A.  The Order by Order Calculation (OBO)}
\nobreak
This means that we simply expand the exponential in Eq.~(2.9) order by
order.
$$\eqalignno{{\cal P} \, \exp\Bigl(\int_{-\infty}^Z  dZ_1  \widehat V (Z_1)
\Bigr)
&= 1 + \int_{-\infty}^Z dZ_1 \widehat V (Z_1)
+ \int_{-\infty}^Z
dZ_1 \widehat V (Z_1) \int_{-\infty}^{Z_1}
dZ_2 \widehat V (Z_2)\cr\cr
&\quad + \int_{-\infty}^Z
dZ_1 \widehat V (Z_1) \int_{-\infty}^{Z_1}
dZ_2 \widehat V (Z_2) \int_{-\infty}^{Z_2}
dZ_3 \widehat V (Z_3) \cr
& \quad + \cdots\quad .&(2.16)\cr}$$
This approximation scheme should be accurate when $\widehat V$ acts as a small
number.  This occurs at high enough energies such that wave packet expansion
effects are minimal.
\medskip
\centerline{B.  The Low Energy Expansion (LEE)}
\nobreak
Before considering the low energy expansion in detail, we first obtain
a general result.  We write $\widehat V$ as a sum of two terms,
$$\widehat V (\B,Z) = V_L (\B,Z) + \Delta V_L(\B,Z),\eqno(2.17)$$
with
$$\eqalignno{V_L & \equiv \langle N | \widehat V  | N \rangle ,&(2.18a)\cr
\Delta V_L & = \widehat V - V_L. &(2.18b) \cr}$$
Notice that with the above definitions, $\langle N | \Delta V_L | N \rangle
= 0$.  Further, if $V_L$ and $\widehat V$ are local operators in nuclear space,
 as in Eq.~(2.15b), then
$$\left[V_L, \Delta V_L \right] = 0\,. \eqno(2.19)$$
Let us suppress the transverse nuclear coordinate, $\B$,
since it is the same for all terms.  Then, in general,
$$\eqalignno{{\cal P}& e^{\int_{-\infty}^Z dZ_1 \Bigl(V_L(Z_1)+\Delta V_L(Z_1)
\Bigr)} = e^{\int_{-\infty}^Z dZ_1 V_L (Z_1)}\cr\cr
&\qquad\quad + \int_{-\infty}^Z dZ_2
e^{\int_{Z_2}^Z dZ_1 V_L (Z_1)} \Delta V_L(Z_2) e^{\int_{-\infty}^{Z_2} dZ_1
V_L (Z_1)}\cr\cr
&\qquad\quad + \int_{-\infty}^Z dZ_1 \int_{-\infty}^{Z_1} dZ_2
e^{\int_{Z_1}^Z
dx V_L (x)} \Delta V_L(Z_1) e^{\int_{Z_2}^{Z_1} dx V_L (x)} \Delta V_L(Z_2)
e^{\int_{-\infty}^{Z_2} dx V_L (x)}\cr
&\qquad\quad+ \cdots\qquad .&(2.20)\cr}$$
The low energy expansion is obtained by using
Eq.~(2.19) to simplify the above equation:
$$\eqalignno{{\cal P} e^{\int_{-\infty}^Z dZ_1 \Bigl(V_L(Z_1)+\Delta V_L(Z_1)
\Bigr)}&=          e^{\lambda (Z)}
\biggl(1+ \int_{-\infty}^Z dZ_1 \Delta V_L(Z_1)\cr\cr
&\quad + \int_{-\infty}^Z dZ_1 \Delta
V_L(Z_1) \int_{-\infty}^{Z_1} dZ_2 \Delta V_L(Z_2) +
\cdots\biggr),&(2.21)\cr}$$
where
$$\lambda (\vec B, Z) \equiv \int^Z_{- \infty} \, dZ_1 \, V_L \, (\vec B, Z_1).
\eqno(2.22)$$
This is called a low energy expansion, since at low energies the wave
packet size increases and $\V
\approx V_L$. Then, $\Delta V_L$ is small and one recovers
 the results of low
energy nuclear physics.
\medskip
\centerline{C.  The High Energy Expansion (HEE)}
\nobreak
Here we also
define a large $V_H$
and small  $\Delta V_H$  potentials and treat $\Delta V_H$ with
perturbation theory.  The separation of terms  is now based on
$\Delta V_H$ acting as small at high energies. Thus
we define
$$\eqalignno{\widehat V & = V_H + \Delta V_H, &(2.23a) \cr
\Delta V_H & = e^{i \phat Z} \V \, e^{-i \phat Z} \, - \widehat
V. &(2.23b) \cr}$$
This approximation is expected to be valid at high energies
because, in the closure limit, all
momenta are equal and $\Delta V_H = 0 $.
The operators $V_H$ and $\Delta V_H$ do not commute, so we must use the more
difficult
form, Eq.~(2.20), of the path ordered exponential in Eqs.~(2.9) and (2.14).
\medskip
\centerline{D.  The Exponential Approximation (EA)}

This is an approximation first made by Miller and Jennings
which turns out to be useful to examine. These authors make the
following approximation:
$$\eqalignno{\sum_n |n \rangle \langle n | e^{i\phat Z} {\cal P} \, & \exp
\int^Z_
{- \infty} \, \widehat V \, (Z_1) dZ_1 |N \rangle\cr
& \approx e^{i\phat Z} \exp \left[ \sum_n \, |n \rangle \langle n |
 \int^Z_{- \infty} dZ_1 \,
\widehat V \, (Z_1) \right]|N \rangle.&(2.24) \cr}$$
At high energies $\V$ is small and the exponential is well approximated
by its first-order expansion. The approximation is also accurate at low
energies because
the exponential accounts for an exact treatment of the usual distorted wave
function.
\bigskip
\centerline{III.  MODEL EVALUATIONS}
\nobreak
The aim here is to determine the regions of accuracy for the various
approximations of Sect. II.  We do this by using tractable models for the
baryon space interaction operator $\widehat V$ and initial wave packet.
The baryonic states are chosen as those of a two dimensional (transverse)
harmonic oscillator that binds
 two quarks.  We take $\hbar\omega = 500 \, MeV$.  The interaction is
specified as
$$f (b^2) = b^2/b^2_H, \eqno(3.1)$$
 in the
notation of Eqs.~(2.11).  This is a simple function with the correct
general properties.  A more elaborate form is studied in Sect. IV.
Now consider the initial wave packet.  We assume that the hard scattering,
represented by $T_H$, creates a wavepacket of zero-size.
Such an object
is called a Point-Like configuration, or PLC. Writing down only the internal
(quark) degrees of freedom, we have
$$\eqalignno{\langle N | T_H (Q^2) & = \sum_n \, F_{N,n} \, (Q^2) \langle n |,
&(3.2a)\cr
& = C (Q^2) \langle \vec b = 0 | \,. &(3.2b) \cr}$$
So, $F_{N,n} (Q^2) = \langle \vec b = 0 | n \rangle C (Q^2)$.  For the two
dimensional oscillator, the matrix element $\langle \vec b = 0 | n \rangle$
has the same value for all $n$,
$\left({1\over{b_H \sqrt{\pi}}}\right)$, and all form factors are equal in
this model.

We now turn to the evaluation of Eq.~(2.14).  Inserting a complete set of
oscillator states gives
$${\cal M}_\alpha  = \int \, d^3 R \, e^{-iqZ} \Phi_\alpha (\vec R) \,
\sum^\infty_{m = 0} \, F_{N,2m} (Q^2)\,  e^{ip_{2m}Z} \langle 2m | {\cal P}
\exp \left(\int^Z_{- \infty} dZ_1 \widehat V (\vec B , Z_1) \right) | N \rangle
\eqno(3.3)$$
where only the isotropic state $n = 2m$ need be kept.  The quantum number $m$
represents the number of nodes in the $b$-space baryon wavefunctions. With
these preliminaries out of the way, we
turn to the evaluation of the different approximations.
\medskip
\centerline{A.  Order by Order Calculation}
\nobreak
The order by order calculation is to evaluate Eq.~(2.16).  The result is
$${\cal M}_\alpha = F_{NN} \, \int d^3 R \, \Phi_\alpha (\vec R) e^{-iqZ} \Psi
(\vec R)\,, \eqno(3.4)$$
where
$$\Psi (\vec R) = e^{ipz} \sum_{n = 0} \Psi_n^{OBO} (\vec R)\,, \eqno(3.5)$$
and the index $n$ is the order of the term in $\widehat V$.
 We find
$$\eqalignno{\Psi^{OBO}_0 (\vec R) &= 1,&(3.6)\cr\cr
         \Psi^{OBO}_1 (\vec R) &= -\so2\iout Z{Z_1} \left(1 - e^{i(p_2-p)
(Z-Z_1)} \right),&(3.7)\cr\cr
         \Psi^{OBO}_2 (\vec R) &= \left(\so2\right)^2 \iout Z{Z_1}
\iout {Z_1}{Z_2}
\Bigl[ 1 -  e^{i(p_2-p)(Z-Z_1)}\cr
                         &\quad  + e^{i(p_2-p)(Z_1-Z_2)}
-3e^{i(p_2-p)(Z-Z_2)}+2e^{i(p_2-p)(Z_1-Z_2) + i(p_4-p)(Z-Z_1)}\Bigr],&(3.8)\cr
\cr
         \Psi^{OBO}_3 (\vec R) &=-\left({{\sigma}\over 2}\right)^3
\int_{-\infty}
^Z dZ_1 \rho (\B,Z_1) \int_{-\infty}^{Z_1} dZ_2 \rho (\B,Z_2)
\int_{-\infty}^{Z_2} dZ_3 \rho (\B,Z_3)\cr
       &\quad \times \Bigl[1-e^{i(p_2-p)(Z-Z_1)}+e^{i(p_2-p)(Z_1-Z_2)}
-3e^{i(p_2-p)(Z-Z_2)}\cr
       &\quad +2e^{i(p_2-p)(Z_1-Z_2)+i(p_4-p)(Z-Z_1)}+e^{i(p_2-p)(Z_2-Z_3)}
 -e^{i(p_2-p)(Z-Z_1+Z_2-Z_3)}\cr
       &\quad +3e^{i(p_2-p)(Z_1-Z_3)}-9e^{i(p_2-p)(Z-Z_3)}
 +6e^{i(p_2-p)(Z_1-Z_3)+i(p_4-p)(Z-Z_1)}\cr
       &\quad -4e^{i(p_2-p)(Z-Z_3)+i(p_4-p_2)(Z_1-Z_2)}
+10e^{i(p_4-p)(Z-Z_2)+i(p_2-p)(Z_2-Z_3)}\cr
       &\quad
-6e^{i(p_6-p)(Z-Z_1)+i(p_4-p)(Z_1-Z_2)+i(p_2-p)(Z_2-Z_3)}\Bigr]\quad,&(3.9)
\cr}$$
\noindent where $p$ is the momentum of the outgoing nucleon and
$p_{2m}=p^2+M^2-M_{2m}^2$.
Note that the above
wavefunctions, $\Psi_1^{OBO}$, $\Psi_2^{OBO}$, and $\Psi_3^{OBO}$ all vanish
in the closure limit.  The order by order curves
are shown in Fig.~1.  The curves labelled by $OBO_n$ are defined by using
$\Psi^{OBO_n} \equiv \sum \limits^n_{m = 0} \, \Psi^{OBO}_m$.

The distorted wave Born approximation
DWBA is obtained by using a scattering wave function
approximated by the neglect of excited baryon states.
Then,
since there are no quark space operators, the exact answer is obtained by
simply exponentiating the first order result without excited states,
$\Psi = e^{ipZ} \left(1-{\sigma\over 2} \int_{-\infty}^Z dZ_1 \rho(\B,Z_1)
\right)$.
That is,
$$\Psi_{DWBA} = e^{ipZ} e^{-\so2\iout Z{Z_1}} \equiv e^{ipZ} e^{\lambda(\B,Z)},
\eqno(3.10)$$
\noindent
where
$$\lambda (\B,Z) = -\so2\iout Z{Z_1}.\eqno(3.11)$$
The name DWBA arises from the notion that the plane wave factor of Eq.~(3.10)
is ``distorted" by a factor, here an exponential damping factor.
The DWBA curve is denoted by the straight solid line on all graphs.
We see that for $-Q^2 \simeq 30 GeV^2$ the second and third order
calculations are converging.  At lower energies, there is poor convergence.
 This is expected, however, since at low energies we expect there to be
many (more than two or three) scatterings.
\medskip
\centerline{B.  The Low Energy Expansion (LEE)}
\nobreak
\noindent Now let us proceed to describe the low energy expansion. With
the present $b^2$ scattering operator, the operators of Eqs.~(2.17) and (2.18)
become
$$\eqalignno{V_L(\B,Z) &= -\so2 \rho (\B,Z),&(3.12)\cr\cr
\Delta V_L(\B,Z) &= -\so2\rho (\B,Z) e^{-i\phat Z}\Delta b^2
e^{i\phat Z}\,.&(3.13)\cr}$$
\noindent where $\Delta b^2 = {b^2 \over b^2_H} - 1$.
Here too, we have gone to third order in $\Delta V_L$.  Using previous
notation, we obtain
$$\eqalignno{\Psi^{LEE}_0 (\vec R) &= e^{\lambda (Z)},&(3.14)\cr\cr
         \Psi^{LEE}_1 (\vec R) &= \so2 e^{\lambda (Z)}
 \iout Z{Z_1} e^{i(p_2-p)(Z-Z_1)},&(3.15)\cr\cr
         \Psi^{LEE}_2 (\vec R) &= \Bigl(\so2\Bigr)^2 e^{\lambda (Z)}
\iout Z{Z_1}
\iout {Z_1}{Z_2}\Bigl[ e^{i(p_2-p)(Z_1-Z_2)}\cr
    &\quad  -
2e^{i(p_2-p)(Z-Z_2)}+2e^{i(p_2-p)(Z_1-Z_2)+i(p_4-p)(Z-Z_1)}\Bigr],
&(3.16)\cr\cr
         \Psi^{LEE}_3 (\vec R) &=\Bigl(\so2\Bigr)^3  e^{\lambda (Z)}
\iout Z{Z_1}
\iout {Z_1}{Z_2} \iout {Z_2}{Z_3}\cr
 &\quad \times \Bigl[ e^{i(p_2-p)(Z-Z_1+Z_2-Z_3)}-2e^{i(p_2-p)(Z_1-Z_3)} +
4e^{i(p_2-p)(Z-Z_3)}\cr
&\quad -4e^{i(p_4-p)(Z-Z_1)+i(p_2-p)(Z_1-Z_3)}
 -8e^{i(p_4-p)(Z-Z_2)+i(p_2-p)(Z_2-Z_3)}\cr
&\quad +4e^{i(p_2-p)(Z-Z_1+Z_2-Z_3) +i(p_4-p)(Z_1-Z_2)}\cr
&\quad
 +6e^{i(p_6-p)(Z-Z_1)+i(p_4-p)(Z_1-Z_2)+i(p_2-p)(Z_2-Z_3)}\Bigr].&(3.17)\cr}$$
\noindent In the closure limit, the wavefunctions given by Eqs.~(3.14 - 3.17)
do not vanish but rather conspire to form the first three terms in
the expansion of $e^{-\lambda(Z)}$, times $e^{+\lambda(Z)}$.
Therefore, to all
orders, the full wavefunction converges to the Born wavefunction, corresponding
to transparency.  Plots of cross sections using the above distorted waves
appear on Fig. 2.  The curves labelled by $LEE_n$ are defined by using
$\Psi^{LEE_n} \equiv \sum \limits^n_{m=0} \, \Psi^{LEE}_m$.
The DWBA curve is $\Psi = e^{ipZ} \Psi_0^{LEE}$.
As before, the Born term is given by $\Psi(\vec R) = e^{ipZ}$.
 The fact that the $LEE$ does not join smoothly up to the DWBA is
evidence that the $m = 1$ ``Roper"  has influence below threshhold.
\medskip
\centerline{C.  The High Energy Expansion (HEE)}
\nobreak
In the present model, the evaluation of Eqs.~(2.23a) and (2.23b) leads to
$$\eqalignno{V_H(\B,Z) &= -\so2 \rho (\B,Z) {b^2\over b^2_H},&
(3.18)\cr
\Delta V_H(\B,Z) &= -\so2\rho (\B,Z) \left[e^{-i\phat Z} {b^2\over b^2_H}
e^{i\phat Z} - {b^2\over b^2_H} \right].&(3.19)\cr}$$
Because $V_H$ and $\Delta V_H$ no longer commute, we must use the path ordered
 exponential shown in Eq.~(2.20).
The result is
$$\eqalignno{\Psi_1^{HEE}(\vec R) &= e^{ipZ} \biggl(\sum_m {{x^{m}(Z)}\over
{1-\lambda (Z)}} e^{i(p_{2m}-p)Z} + \sum_{m,n} x^{n+1}(Z) e^{i(p_{2m}-p)Z}
\langle 2m \mid {b^2 \over b^2_H}  \mid 2n \rangle\cr
&\qquad - \so2 \int_{-\infty}^Z dZ_1 {{\rho(Z_1)}\over {1-
\lambda (Z_1)}} \sum_{m,n,l} e^{i(p_{2m}-p)Z} e^{i(p_{2n}-p_{2l})Z_1}
 \, x^{n}(Z_1)\cr
&\qquad\qquad \times \langle 2m \mid e^{\lambar (Z, Z_1) {b^2 \over b^2_H}}
\mid 2l \rangle \langle 2l \mid {b^2 \over b^2_H} \mid 2n\rangle
\biggr),&(3.20)
\cr}$$
\noindent where $\lambda (Z)$ is defined in Eq.~(3.11) above,
$$\lambar (Z, Z_1) = -\so2 \int_{Z_1}^Z dZ_2 \rho (Z_2),
\eqno(3.21)$$
$$ x(Z_1)   = {{-\lambda (Z_1)}\over {1-\lambda (Z_1)}},\eqno(3.22)$$
and
$$\langle 2m \mid e^{\lambar {b^2 \over b^2_H}} \mid 2l \rangle =
-\lambar \biggr(
{1\over {\lambar (-1+\lambar )}}\biggr)^{l+m+1} \,\sum_{i=0}^{min(l,m)}
{m\choose i}
{l\choose i} \Bigl( \lambar^2 \Bigr)^{l+m-i}.\eqno(3.23)$$
The result (3.23) is obtained from
$$\langle b | 2m \rangle = {1\over {b_H \sqrt{\pi}}}\sum_{i=0}^m {{(-1)^i}\over
 {i!}} {m\choose i} e^{-{b^2}\over {2 b_H^2}}\left({{b^2}\over {b_H^2}}
\right)^i\,.
\eqno(3.24)$$
Although it is not immediately obvious, $\Psi_1^{HEE} (\vec R)$, given by
Eq.~(3.20), in the
closure limit, does indeed approach unity and color transparency is
obtained.
Notice that we
have suppressed the transverse nuclear
distance, $\B$, in $\lambda$, $x$,  and $\Psi$ for clarity.

At this stage, a further approximation suggests itself.  At very high
energies, considering the form of $\phat$ in Eq.~(2.3), it seems reasonable to
replace
$$p_{2m} - p \rightarrow m(p_2-p).\eqno(3.25)$$
This is what we call the ``Equal Spaced Momentum'' (ESP)
replacement.
With the above replacement, we can now do the sum on $l$ in Eq.~(3.20) with
the result that
$$\eqalignno{ \Psi_1^{HEE}(\vec R) &= e^{ipZ} \biggl( {1\over {(1-\lambda (Z))
(1-y(Z,Z))}}
+{{\Bigl(x(Z) - y(Z,Z) \Bigr) \Bigl(1-x(Z) \Bigr)}\over
{ \Bigl(1-y(Z,Z) \Bigr) ^2}}\cr
&\quad - \so2 \int_{-\infty}^Z dZ_1 {{\rho(Z_1)}\over {1-\lambda
(Z_1)}} \sum_{m,n} e^{i(p_2-p)mZ} \, x^{n}(Z_1) \qquad\qquad \times\cr
&\quad \Bigl((2n+1)\langle 2m \mid e^{\lambar (Z,Z_1) {b^2 \over b^2_H}}
\mid 2n \rangle\cr
&\quad
-(n+1) e^{i(p-p_2)Z_1} \langle 2m \mid e^{\lambar (Z, Z_1)
{b^2 \over b^2_H} } \mid 2n+2 \rangle\cr
&\quad - n e^{i(p_2-p)Z_1} \langle 2m \mid
e^{\lambar (Z, Z_1) {b^2 \over b^2_H}} \mid 2n-2 \rangle\Bigr)
 \biggr).&(3.26)\cr}$$
The new function $y$ is given by
$$y(Z,Z_1) = x(Z_1) e^{i(p_2-p)Z}.\eqno(3.27)$$
Intermediate states
corresponding to $m = n = 9$ must be included
to accurately obtain numerical results for expressions (3.20) and (3.26).
On Fig.~3, we see this first order term of the $HEE$ (solid) compared with
the $OBO_3$ (dotdashed).  As expected, the two curves agree at
higher energies. The ESP replacement is very accurate at all but the very
lowest energies.
\medskip
\centerline{D.  The Exponential Approximation (EA)}
\nobreak
If one looks at only the
first order term in Section IIIA, we see that the only thing which has changed
between this wave function calculation and the traditional Glauber
calculation is that the cross section $\sigma$ has been replaced by an
effective cross section $\sigma (Z,Z_1) \equiv \sigma (1-
e^{i(p_2-p)(Z-Z_1)})$.  Then, it is quite tempting to simply exponentiate
this first order result.  This so-called exponential approximation  is
not correct, since only the second excited state enters
whereas we know that higher states enter in higher orders.  However,
because we have learned that the ESP replacement is a safe one, we can cast
all higher excitations in terms of excitations of the second state.  In
fact, if we examine the difference $\Delta_2$ between the exact second order
OBO
calculation and the second order term of the EA, we find that,
$$\Delta_2 = \Bigl( \so2 \Bigr)^2 \iout Z{Z_1} \iout {Z_1}{Z_2}
e^{i(p_2-p)(Z_1-Z_2)}
\Bigl(1-e^{2i(p_2-p)(Z-Z_1)}\Bigr)^2,\eqno(3.28)$$
where we have used the ESP replacement.  Therefore, we see why the
EA works as well as it does.  The exponentiation of the first OBO contains
many more of the higher order terms than one would naively think.
Thus,
at high energies, the difference between the EA and the true answer is
rapidly suppressed, as shown above.  The EA
(dotted) is shown
along with the first order OBO calculation on Fig. 4.
\medskip
\centerline{E.  Summary of Approximation Schemes}
\nobreak
We have presented many different approximations and graphs.
Here, we compare and contrast the results of the above calculations.  In
particular, we want to see which approximation method or methods work
best.  To do this, we study the convergence properties among all curves.
Looking at the Figures we see that:

\item{1.}  The curves $OBO_2$ and $OBO_3$ converge at about $-Q^2 \approx 30
GeV^2$.
This leads us to believe that above  $-Q^2 \approx 30 GeV^2$, the second/third
order result is the true answer in this model.  At low energy, we see no
convergence.  This is not surprising since at low energies, we expect there to
be many (more than two or three) scatterings.

\item{2.}  The curves $LEE_2$ and $LEE_3$ follow each other closely up until
$-Q^2
\approx 20 GeV^2$.  This indicates that $LEE_3$ is the correct answer at least
until  $-Q^2 \approx 20 GeV^2$, and probably further, depending on how big the
fourth order contribution is.  Comparing Figs. 1 and 2, we see that $LEE_3$
and $OBO_{2,3}$ line up at
about  $-Q^2 \approx 26 GeV^2$.  This gives us confidence that $LEE_3$ is the
correct answer for $-Q^2 \approx 26 - 35 GeV^2$.  Therefore, we can conclude
that $LEE_3$ is the correct answer all the way up to  $-Q^2 \approx 35 GeV^2$.
 Of course, at higher energies, we expect $LEE_3$ to be inaccurate, as did
$LEE_1$, and higher order terms would be needed.  Not bad, however, for a low
energy expansion.

\item{3.}  Fig. 3 shows the results of the high energy expansion including
the ESP
replacement.  Also shown for comparison is the $OBO_3$.  The two curves are
shown to line up for energies higher than $-Q^@ \approx 20 \, GeV^@$
indicating that the true answer in this model is well approximated by the
$HEE$.

\item{4.}  Fig. 4 shows the exponential approximation and the $OBO_1$ for
comparison.
The $EA$ matches the $HEE$ all the way to  $-Q^2 \approx 6 GeV^2$, indicating
that the $EA$ is a simple and fast way to calculate the distorted wave for
high energies.

\item{5.}  The most successful of these approximations are the $LEE_3$, the
$HEE$ and the $EA$.  They are compared in Fig. 5.
\bigskip
\centerline{IV.  MORE GENERAL INTERACTIONS}
\nobreak
The advantage of the present formalism is its applicability to general forms
of the interaction, $f (b^2)$ of Eqs.~(2.11).  The tests of Sect. III were
performed with $f(b^2) = b^2/b^2_H$.  Here we use the form of Nikolaev and
Zakharov~[15], which is well approximated~[16] by
$$f (b^2) = {{1 + \gamma}\over {\gamma}}
\, \left(1 - e^{- \gamma b^2/b^2_H} \right)\,.
\eqno(4.1)$$
This is derived for $ q \bar q$ color singlet wave packets, but we use it here
as a general representative of Eqs.~(2.11).  The parameter $\gamma$ = 0.762.

We use the LEE to determine the cross sections predicted by this
interaction.
Then the result Eq.~(2.18a) is unchanged and the new form of $\Delta V_L$ is
$$\Delta V_L = - {\so2} \rho (R) {1 \over \gamma} \, \left[1 - (\gamma
+ 1) e^{- \gamma b^2/b^2_H} \right]. \eqno(4.2)$$
A straightforward application leads to the results of Fig. 6.  At high
energies the $OBO_2$ and $LEE_2$ approximations are in agreement.  At low
energies the results of $LEE_2$ agree with those of the EA.  Thus the
$LEE_2$ provides a good representation of the correct answer for this model.
We see that the results are qualitatively similar to those of Ref.~[13], but
the predicted ratios of $\sigma/\sigma_{B}$ are reduced by about 10\%.
\bigskip
\centerline{V.  FINITE SIZE EFFECTS}
\nobreak
So far we have assumed that
a $PLC$ was made in the hard scattering.  Another consequence of this
assumption, also used, is that the elastic and inelastic form factors,
$F_{NN}$ and $F_{2m,N}$, are equal.  Yet another way to restate
Eqs.~(3.2) is to assert that  $T_H(\vec
b) \propto \delta^{(2)} (\vec b)$, where $\vec b$ is the transverse distance
between the
two quarks in the struck nucleon.  We now examine the effects of including
a nonzero size for the $PLC$. In this section, we use the simpler form
($\sim b^2$) for
the wavepacket-nucleon interaction.

	We have incorporated the effects of finite size in two different ways.
In the first case, we examine the effect which letting $T_H(\vec b)$ have a
nonzero range has on the limit of ${ {\sigma}\over {\sigma_{B}}}$.  Here,
we assume
$$T_H(\vec b)= {1\over {\lam \pi}} e^{-{{b^2}\over {\lam}}}.\eqno(5.1)$$
It is then simple to calculate the form factors.  In fact, a
calculation similar to that for deriving Eq.~(3.24) enables us to calculate
all of the form factors at once.  The result is
$$F_{2m,N} = {1\over {\pi \bigl(\lam + b_H^2\bigr)}} \biggl( {{b_H^2}\over
{\lam + b_H^2}}\biggr)^m,\eqno(5.2)$$
where we take $b_H = 1fm$.  In order to illustrate the effects of finite
size, we examine the $OBO_1$ case for simplicity.  In the case of unequal form
factors, the distorted $OBO_1$ wave becomes
$$\Psi^{OBO_1} (\vec R) =1-\so2 \iout Z{Z_1}\Bigl(1 - {{F_{2,N}}
\over {F_{N,N}}} e^{i(p_2-p)(Z-Z_1)}
\Bigr).  \eqno(5.3)$$
 This is to be compared with Eqs.~(3.6) and (3.7).
Fig. 7 shows the
ratio of $\sigma$ to $\sigma_{B}$ for a very large value of $-Q^2$.  As
$\lam$ approaches zero, $T_H(\vec b)$ approaches
a delta function, and the transparency is unity.  As $\lam$ goes to
infinity, the ratio of the form factors goes to zero, and the first order
term of the standard DWBA results.  This shows that the predicted cross
sections are controlled by the existence of a PLC.

The immediate question is, then, what is $\lam$?  Above, it is an
arbitrary parameter.  But, if we believe that a small sized configuration is
produced in the hard interaction, then $\lam$ should be related to the
momentum transfer of the incoming
photon to the ejectile.  Suppose we have a nucleon of three quarks which
absorbs a photon of three momentum $\vec q$.  Each quark, then, gets
approximately $q/3$ of the photon's momentum.  So, we use a form~[17] for
the hard scattering operator
$$T_H(\vec b)= e^{{-{b_H q}\over 3}}\,.\eqno(5.4)$$
This form is suggested by caricatures of pQCD calculations; see Eq.~(1) of
Ref.~[13].  With Eq.~(5.4) in hand,
the form factors can be obtained in closed form.  The
results are
$$\eqalignno{F_{NN} &= 1 - e^{x^2} \sqrt{\pi} \, x \, {\rm Erfc}(x),&(5.5)\cr
F_{2,N} &=- x^2 + {1\over 2} e^{x^2} \sqrt{\pi} \, x \, {\rm Erfc} (x)
+ e^{x^2} \sqrt{\pi} \, x^3 \, {\rm Erfc}(x).&(5.6)\cr}$$
where $x ={1\over 2} {{q b_H}\over 3}$  and $\rm{Erfc}(x)$ is the
complementary error function.  We use Eq.~(5.3) to calculate the cross
section .
 Fig. 8 shows the finite
sized result obtained using Eqs.~(5.5) and (5.6), labelled by $OBO_{FS}$,
compared
with the zero-size $OBO_1$.  We see that including finite size, via Eq.~(5.4)
has a very
small effect.
\bigskip
\centerline{VI.  $N^*$ PRODUCTION}
\nobreak
We can also apply our
machinery to examine $N^*$ production.
Very little changes in the formalism if a PLC is formed and if
the $b^2$ form of the wave-packet nucleon interaction is used.

Let us first consider the DWBA for $N^*$'s.  The $N^*$
experiences its own optical potential as it moves through
the nucleus.   We can obtain the
result for the distorted wave by neglecting states other than the $N^*$, e.g.
make $\U$ diagonal in the quark space.  The result is
$$\Psi_{DWBA}(\vec R)=e^{ip_2Z} e^{-{{3\sigma}\over 2}\int_{-\infty}^Z dZ_1
\rho(\B,Z_1)}\equiv e^{ip_2Z}e^{\lambda^*(\B,Z)},\eqno(6.1)$$
where
$p_2$ the momentum of the outgoing excited state is close to the
photon momentum $\vec q$. Note  the
appearance of the factor 3 in the exponent of Eq.~(6.1).
This factor has a large effect, causing terms in which the $N^*$ decays to
be more important than those in which the $N^*$ propagates.

Using previous notation, we have calculated $OBO_1^*$, $OBO_2^*$, $LEE_1^*$,
$LEE_2^*$, and $EA^*$.  The superscripts serve to remind us that we are
calculating $N^*$ terms.

We first show the $OBO^*$ results.  Recall, $\Psi(\vec R)=e^{ip_2Z} \bigl[
\Psi_0^{OBO^*}(\vec R) + \Psi_1^{OBO^*}(\vec R) +\Psi_2^{OBO^*}(\vec R) +
\cdots\bigr]$.
$$\eqalignno{\Psi_0^{OBO^*}(\vec R) &= 1,&(6.2)\cr\cr
\Psi_1^{OBO^*}(\vec R) &= -\so2\iout Z{Z_1} \bigl( 3 - e^{i(p_0-p_2)(Z-Z_1)}
-2e^{i(p_4-p_2)(Z-Z_1)}\bigr),&(6.3)\cr\cr
\Psi_2^{OBO^*}(\vec R) &=\bigl(\so2\bigr)^2\iout Z{Z_1}\iout {Z_1}{Z_2}
\bigl( 9 \cr
&\quad +4e^{i(p_4-p_2)(Z_1-Z_2)} + e^{i(p_0-p_2)(Z_1-Z_2)}-e^{i(p_0-p_2)(Z-
Z_2)}\cr
&\quad -3e^{i(p_0-p_2)(Z-Z_1)}-6e^{i(p_4-p_2)(Z-Z_1)}-10e^{i(p_4-p_2)(Z-Z_2
)}\cr
&\quad+6e^{i(p_6-p_4)(Z-Z_1) + i(p_4-p_2)(Z-Z_2)}\bigr).&(6.4)\cr}$$
\noindent The $OBO^*$ curves are shown on Fig. 9 with the same notation
as before.  The lower solid line is the full distorted wave Born
approximation, denoted by $DWBA^*$; the upper solid line is
the cross section calculated using only the first order term of the DWBA
and is labelled $DWBA_1^*$.
The difference between those two curves is much bigger than in the nucleon
case (the relevant lines there would correspond to $DWBA$ = .54,
$DWBA_1$ = .48)
This difference is simply due to the presence of the extra factor of
3 in Eq.~(6.1).  The resonance-like bump at lower energies is a false
effect.  Because the first order term starts out so big, which we see is
because
of the large value produced by the first order DWBA, the tendency of the
phases in Eqs.~(6.2--6.4) is to increase the cross section at the lower
energies.  It is not surprising that
the order by order calculation fails here.  Indeed, we learned from our
experience with the nucleon that the $OBO$ curves were unreliable at low
energies.  The fact that the exponent of the DWBA is large in magnitude only
accentuates that failure.  On these grounds, we expect the $LEE^*$ and the
$EA^*$ curves, which have many more higher order effects, to do much better.

Of course, the $EA^*$ is calculated by simply exponentiating the $OBO_1^*$,
Eq.~(6.3).  The $LEE^*$ results for the distorted waves are shown below.
$$\eqalignno{\Psi_0^{LEE^*}(\vec R) &= e^{\lambda(\B,Z)},&(6.5)\cr\cr
\Psi_1^{LEE^*}(\vec R) &=-\so2 e^{\lambda(\B,Z)}\iout Z{Z_1} \biggl(2-e^{i(p_0-
p_2)(Z-Z_1)}\cr
& \qquad \qquad  - 2e^{i(p_4-p_2)(Z-Z_1)}\biggr),&(6.6)\cr\cr
\Psi_2^{LEE^*}(\vec R) &=\bigl(\so2\bigr)^2 e^{\lambda(\B,Z)}\iout Z{Z_1}
\iout {Z_1}{Z_2} \biggl( 4 +4e^{i(p_4-p_2)(Z_1-Z_2)}\cr
&\quad  + e^{i(p_0-p_2)(Z_1-Z_2)}-2e^{i(p_0-p_2)(Z-Z_1)}-4e^{i(p_4-p_2)(Z-Z_1)}
\cr
&\quad -8e^{i(p_4-p_2)(Z-Z_2)}
+6e^{i(p_6-p_4)(Z-Z_1) + i(p_4-p_2)(Z-Z_2)}\biggr).&(6.7)\cr}$$
Here, $\lambda(\B,Z)$ is defined in Eq.~(3.11).  The $LEE_1^*$, $LEE_2^*$,
and $EA^*$ are shown on Fig. 10 as the dotted,
dashed, and solid lines, respectively.  As expected, there is good
agreement among all three curves, especially below $- Q^2 \sim $ 20 $GeV^2$.

It is curious that the $LEE^*$ works as well as it does since now
the 0th order term does not reproduce the $DWBA^*$.  Therefore, one
might think that instead of defining ${b^2 \over b^2_H} = 1 + \Delta b^2$ it
would be
more appropriate to define the operator $\overline{\Delta b^2}$ by the
assignment ${b^2 \over b^2_H} = 3 + \overline{\Delta b^2}$ and then use
Eqs.~(3.12~--~3.13).  Calculations using this ``bar" method show that
the that convergence is better in the $LEE^*$ scheme than in the
$\overline{LEE^*}$. This is due to the preference, noted above,
that $N^*$ decay is preferred to $N^*$ propagation.

Note that in this $N^*$ work, we have taken the ejected wavepacket to
be in a PLC.
Qualitative changes in the predictions could occur if this is assumption is
not correct~[11].

We now summarize our results for the $N^*$ case.

\item{1.}  The curves $OBO_1^*$ and $OBO_2^*$ do not really start to converge
well
at the energies shown.  However, starting at about $-Q^2 \approx 16 GeV^2$,
the two curves are not far apart.

\item{2.}  The curves $LEE_1^*$ and $LEE_2^*$ show good agreement for energies
less than $-Q^2 \approx 20 GeV^2$.  Above that, the $LEE_1^*$ is not
accurate.
The fact that the $LEE_2^*$ agrees so well with the $EA^*$ starting
from $-Q^2 \approx 8 GeV^2$ suggests that the correct answer is given by
these two curves above this energy.

\item{3.}  Below  $-Q^2 \approx 8 GeV^2$ the $\overline{LEE_2^*}$ agrees
extremely
well with the $EA^*$.  This indicates that these two curves give the
correct answer for this energy range.  Also, this leads us to the
conclusion that the shoulder which appears at low energies in the $LEE_2^*$
at low energies is not real.

\item{4.}  Our experience in the
nucleon sector taught us that the $LEE$ and the $EA$ were the best
approximations.  This conclusion remains true in the case of $N^*$'s too.

\bigskip
\centerline{VII.  SUMMARY}
\nobreak
We have derived  approximation schemes that are accurate at low (LEE,
   Sect. IIB) and
 high (HEE, Sect. IIC) energies. Their regimes of accuracy overlap, so that
 methods for calculations at any energy exist. The applications here
use simplified interactions and baryon wavefunctions.  The methods
are more general and allow calculations using more realistic interactions
and wavefunctions.

\bigskip
\centerline{ACKNOWLEDGMENTS}

We thank the  DOE for partial support of this work. We thank B. K. Jennings
for useful discussions and K. Nikolaev
for raising questions about Ref.~[13] that caused us to undertake this
investigation.
\vfill\eject
\centerline{\bf REFERENCES}
\item{[1]} A.H.~Mueller, ``Proceedings of Seventeenth Rencontre de
    Moriond", Moriond, 1982 ed. J Tran Thanh Van (Editions Frontieres,
Gif-sur-Yvette, France, 1982)p13.
\item{[2]} S.J.~Brodsky in Proceedings of the Thirteenth
intl Symposium
on Multiparticle Dynamics, ed. W.~Kittel, W.~Metzger and A.~Stergiou (World
Scientific, Singapore 1982,) p963.
\item{[3]}A.S. Carroll et al Phys Rev Lett 61,1698 (1988; S. Heppelmann,
p. 199 in ``Nuclear physics on the Light Cone",ed. by M.B. Johnson and
L.S. Kisslinger , World Scientific (singapore, 1989).
\item{[4]}SLAC Expt. NE-18, R. Milner, Spokesman.
\item{[5]} A.S. Carroll {\it et al.}, BNL expt. 850.
\item{[6]} F.E. Low, Phys. Rev. D12, 163 (1975);
 S. Nussinov Phys. Rev.  Lett 34, 1286 (1975);
J Gunion D Soper Phys Rev D15,2617 (1977).
\item{[7]} L. Frankfurt and M. Strikman, Progress in Particle and Nuclear
Physics, 27,135,(1991);
L. Frankurt and M. Strikman, Phys. Rep. 160, 235 (1988).
\item{[8]} B.Z. Kopeliovich, Sov.J. Part. Nucl. 21, 117 (1990).
\item{[9]} L Frankfurt, G.A. Miller \& M. Strikman, "Color Transparency and
Nuclear Phenomena", to be published Comm. Nuc. Part. Phys. Sept. 1992.
1992 UWA preprint-40427-26-N91.
\item{[10]} H-n Li and G. Sterman,``The perturbative Pion Form Factor with
Sudakov Suppression", 1992 preprint ITP-SB-92-10.
\item{[11]}L. Frankfurt, G.A. Miller \& M. Strikman, Point-like
configurations in Hadrons and Two-body collisions, to be submitted to Nucl.
Phys. A.
\item{[12]}  G.R.~Farrar, H.~Liu, L.L.~Frankfurt \& M.I.~Strikman, Phys.
Rev. Lett. 61 (1988) 686.
\item{[13]} B.K.~Jennings and G.A.~Miller, Phys. Lett. B236, 209 (1990);
 B.K.~Jennings and G.A.~Miller, Phys. Rev. D 44, 692 (1991);
 G.A. Miller and B.K. Jennings p. 25 in "Perspectives in Nuclear Physics
at Intermediate Energies" Ed. S. Boffi, C. Ciofi degli Atti, M. Giannini,
1992 ,World Sci. Press Singapore;
G.A.  Miller, ``Introduction to Color Transparency",
in ``Nucleon resonances and Nucleon Structure",  G.A.
Miller, editor. To be published by World Sci., Singapore (1992).
\item{[14]}  B.Z.~Kopeliovich and B.G.~Zakharov, Phys. Lett. B264 (1991) 434
; Phys Rev D 1992

\item{[15]} N.N.~Nikolaev and B.G.~Zakharov, Z. Phys. C49 (1991) 607
\item{[16]} G.Piller, private communication.
\item{[17]} B.K. Jennings private communication.

\vfill \eject
\centerline{FIGURE CAPTIONS}
\bigskip
\item{Fig. 1.} Order-by-order approximation scheme. Ratios of cross sections
for the $^{12}C(e,e'p)$ reaction. Dotted- OBO$_1$. Dashed-OBO$_2$.
Solid- OBO$_3$.
\medskip
\item{Fig. 2.} The low energy expansion. Dotted-LEE$_1$. Dashed-LEE$_2$.
Solid-LEE$_3$.
\medskip
\item{Fig. 3.} The high energy expansion.
\medskip
\item{Fig. 4.} The exponential approximation.
\medskip
\item{Fig. 5.} Comparing different approximation schemes.
\medskip
\item{Fig. 6.} Use of the interaction of Eq.~(4.1), Ref.~[15].
\medskip
\item{Fig. 7.} The effect of a finite sized wave-packet at very high $Q^2$.
Size $\sim {1\over \Lambda}$.
\medskip
\item{Fig. 8.} Energy dependence of the finite-size effects, $\Lambda\sim Q$.
\medskip
\item{Fig. 9.} Order-by-order approximation for quasi-elastic $N^*$ production.
\medskip
\item{Fig. 10.} Low energy expansion for quasi-elastic $N^*$ production.
\medskip
 \bye